The Computed Microwave Spectrum of the Protonated Fullerene $C_{60}H^+$

*Laszlo Nemes[1*], Jos Oomens[2,3], Vincent J. Esposito[4], Vincent Boudon[5]*

*Alexander G. G. M. Tielens[6]*

1 HUN-REN Research Center for Natural Sciences, Institute of Materials and Environmental Chemistry, 1117 Budapest, Magyar Tudósok Körutja 2, Hungary

2 HFML-FELIX, Toernooiveld 7,6525 ED Nijmegen, The Netherlands

3 Institute for Molecules and Materials, Radboud University, Heyendaalseweg 135, Nijmegen, The Netherlands

4 Schmid College of Science and Technology, Chapman University, 1 University Drive, Orange ,CA 92866, United States

5 Laboratoire Interdisciplinaire Carnot de Bourgogne, UMR 6303 CNRS-Université Bourgogne Europe, 9 Avenue Alain Savary, BP 47 870, F-21078 DIJON Cedex FRANCE

6 Astronomy Department, University of Maryland, College Park, MD 20742

*Correspondence: nemes.laszlo@ttk.mta.hu

**Keywords:**

protonated fullerene $C_{60}$, rotational spectra, rotational simulations, quantum chemical calculations, anharmonic vibrations, radio astronomy

## Abstract

The largest known molecule in space, $C_{60}$ , has been detected in its neutral and cationic form through its vibrational, UV-driven fluorescence emission spectrum and its electronic absorption spectrum, respectively. The detection of several polycyclic aromatic hydrocarbon molecules through their pure rotation spectrum in cold, dense, molecular cloud cores suggests that $C_{60}$ might be present in these environments as well. The low flux of UV pumping photons in molecular cloud cores and the absence of suitably bright background stars, make detection of $C_{60}$ and its cation through the commonly used methods impractical. As $C_{60}$ has no permanent dipole moment, its pure rotational transitions are forbidden and its presence must be inferred from the rotational transitions of $C_{60}$ derivatives with permanent dipole moments. Here, we present a study of the predicted rotational spectrum of protonated $C_{60}$ that has a sizeable permanent dipole moment. Protonation of $C_{60}$ reduces the icosahedral symmetry to $C_s$ and results in a dipole moment of about 3.8 Debye. The resulting $C_{60}H^+$ is a closed shell system with no electron spin. The ground electronic symmetry is A'. The goal of the present calculations is to simulate rotational spectra in radio astronomy frequency ranges. The simulations are based on geometry and electric dipole moment values from harmonic and anharmonic density functional theory (DFT) calculations at the B3LYP level. Using the PGOPHER spectral simulation program, the rotational structure of the spectra at excitation temperatures of 5 and 10 K were computed to guide future laboratory studies and facilitate radio astronomy searches for protonated $C_{60}$ in cold dark molecular clouds.

## 1. Introduction

Following the discovery of the fullerenes[1] in 1985, the macroscopic synthesis of $C_{60}$ by Krätschmer *et al*.[2] and the Nobel prize in Chemistry given to Harold Walter Kroto, Robert Curl and Richard Smalley in 1996, interest in fullerenes quickly increased and resulted in an almost exponential growth of the number of publications. Originally, Kroto was interested in the cosmic occurrence of linear cyanopolyynes, which led to his basic interest in the presence of fullerenes in the interstellar medium. The fundamental question was whether some of the so-called diffuse interstellar bands (DIBs) could also be due to fullerenes. Historical clarification on DIBs started with the first observations by Annie Jump Cannon and E.C. Pickering.[3] The research on DIBs continued in 1922 by the observation of sharp spectral absorption features in astronomical spectra by Mary Lea Heger.[4]

DIBs occur generally as broad diffuse bands in the visible and infrared spectral ranges in the spectra of background stars, where their light traverses relatively low-density material.[5] It was assumed already early on that the DIBs were due to large carbon bearing molecules. To date, there are some 500 known DIBs, but the difficulty in assigning them to specific molecules remains and is illustrated by the fact that so far there is only one definite molecular assignment: the fullerene radical cation $C_{60}^{+}$ having two electronic bands observed in diffuse clouds. After preliminary identification of the $C_{60}^{+}$ DIBs by Foing and Ehrenfreund[6] in 1994, following Fulara et al.[7], these two DIBs were finally identified based on laboratory gas-phase spectra obtained through sophisticated laser spectroscopy methods in the Maier group in 2015.[8] Three weaker DIBs have now also been assigned to $C_{60}^{+}$.[9]

The infrared spectra of many astronomical objects are dominated by broad emission bands due to vibrational relaxation of Polycyclic Aromatic Hydrocarbon (PAH) molecules.[10] These emission features are associated with regions illuminated by strong ultraviolet radiation fields – so-called photodissociation regions[11] – that pump these molecules electronically. The ensuing vibrational cascade produces bright infrared emission that can be detected using *e.g.* space born instruments.[10] In 2010, the vibrational bands of neutral $C_{60}$ and $C_{70}$ were identified in the IR spectrum of the Tc-1 planetary nebula.[12] (See also the interview paper with Jan Cami.[13]) A number of further cosmic $C_{60}$ detections followed in other sources by several authors, *e.g.* also using the Spitzer Space Telescope.[14]

These studies on the presence of $C_{60}$ in space are limited to regions that lie in front of bright background stars – thus allowing electronic absorption spectroscopy – or that are irradiated by strong ultraviolet radiations fields – producing bright infrared emission. However, in recent years, rotational spectroscopy in the GHz range has uncovered unexpectedly high abundances of small aromatic species, in particular indene and derivatives of naphthalene, pyrene and coronene, in dark molecular cloud cores such as TMC1, regions that cannot be probed using visible absorption or IR emission spectroscopy due to the dimness of background stars and the absence of UV radiation.[15,16,17] These aromatic species point toward the importance of gas-phase chemistry routes for the formation of aromatic species inside cold dense molecular cloud cores. It is conceivable that these routes may also form species as large as $C_{60}$. $C_{60}$ and $C_{60}^+$ have no permanent dipole moment and cannot be detected in dark cloud cores through their rotational emission. The best method to infer their presence in these environments is through the rotational emission of $C_{60}$ derivatives with permanent dipole moments, such as protonated $C_{60}$.

In general, the inventory of cosmic fullerenes and PAH molecules is not yet well known and the chemical processes that lead to their formation and destruction are still under investigation. Electronic spectra would be useful to find $C_{60}H^+$ in interstellar spectra. UV transitions were identified in cryogenic neon matrix isolation experiments.[18] However, only broad absorption bands were observed and these are difficult to detect in interstellar spectra. In infrared spectroscopy, molecules may have overlapping IR spectra, challenging their individual detection, as is for instance the case for PAH molecules. In microwave spectroscopy, spectral lines rarely overlap among different molecules as the rotational frequencies are sensitive functions of the rotational constants. Thus, microwave methods provide powerful tools to identify large carbon molecules in space. However, only polar molecules possess rotational spectra and therefore, the basic fullerenes $C_{60}$ and $C_{70}$ cannot be detected by radio astronomy.

Note that most PAH molecules are also apolar. Ionized fullerenes are usually polar and may be accessible by radio methods. An important counter example is $C_{60}^+$, in which ionization leads to a Jahn-Teller effect that reduces the icosahedral symmetry to $D_{5d}$,[19] which is an apolar symmetry group. Thus, $C_{60}^+$ cannot be detected by radio methods. The Jahn-Teller effect in $C_{70}^+$ similarly reduces the $D_{5h}$ symmetry of the neutral molecule to $C_s$ and this ion has a dipole moment of about 1.3 D.[20] In the present work, we report a computational study of the rotational spectrum of protonated $C_{60}$ ($C_{60}H^+$) that has a $C_s$ geometry and an anharmonically computed dipole moment of about 3.8 D. We argue that it could thus be useful as a proxy to detect $C_{60}$ in various cosmic sources. The present calculations of the rotational spectrum are based on harmonic[21] and anharmonic[22] quantum-chemical calculations. Harmonic calculations were used originally for the analysis of the vibrational spectrum of $C_{60}H^+$ by Palotás *et al.*,[21] but the present study reports rotational constants and electrical polarity quantities as well.

**2. Quantum-chemical methods**

The GAUSSIAN 16 suite of programs[23] was employed to run geometry optimization and harmonic frequency calculations of $C_{60}H^+$. The B3LYP functional in combination with the 6-311+G(d,p) basis set was selected because of its favorable performance in predicting the vibrational spectrum.[21] In the calculations presented here the symmetry was constrained to $C_s$ and we verified that all vibrational frequencies are real, so that the corresponding geometry is a true minimum.

To compute the anharmonic spectrum, the geometry of $C_{60}H^+$ was optimized in $C_s$ symmetry, using the B3LYP density functional in combination with the 6-31G basis set. This smaller basis set was used to compensate for the high computational cost of anharmonic frequency calculations for large molecules. The optimization used tight convergence criteria (energy convergence threshold of $1x10^{-12}$) and employed an integration grid consisting of 200 radial shells and 974 angular points per shell (200,974). Following this, the normal modes, harmonic frequencies, and cubic and quartic force constants were computed at the same level of theory. The higher order force constants were computed using numerical differentiation via finite differences of analytical gradients as implemented in Gaussian 16. From there, $2^{nd}$ order vibrational perturbation theory was used to compute the anharmonic vibrational frequencies of $C_{60}H^+$ to obtain the ground-state vibrationally averaged rotational constants as well as the quartic and sextic constants. The size of the computed electric dipole moment is 3.8622 D and the value of the final electronic energy is -2285.928213 Hartree.

## 3. The molecular geometry and spectral parameters of $C_{60}H^+$

Using the ground vibrational state averaged rotational constants and quartic centrifugal constants from anharmonic calculations described in the previous section, the rotational spectra were simulated using the PGOPHER software.[24] The binding of the proton to a carbon atom of $C_{60}$ by a covalent single chemical bond distorts the geometry from its icosahedral symmetry to $C_s$. Figure 1 shows the structure of $C_{60}H^+$ as obtained from harmonic calculations using the B3LYP/6-311+G(d,p) functional constrained to $C_s$ symmetry. The anharmonic calculations were also constrained to $C_s$ geometry and led to a structure with Cartesian atomic coordinates slightly different from the values used for Figure 1, but for structural discussion the harmonic picture suffices.

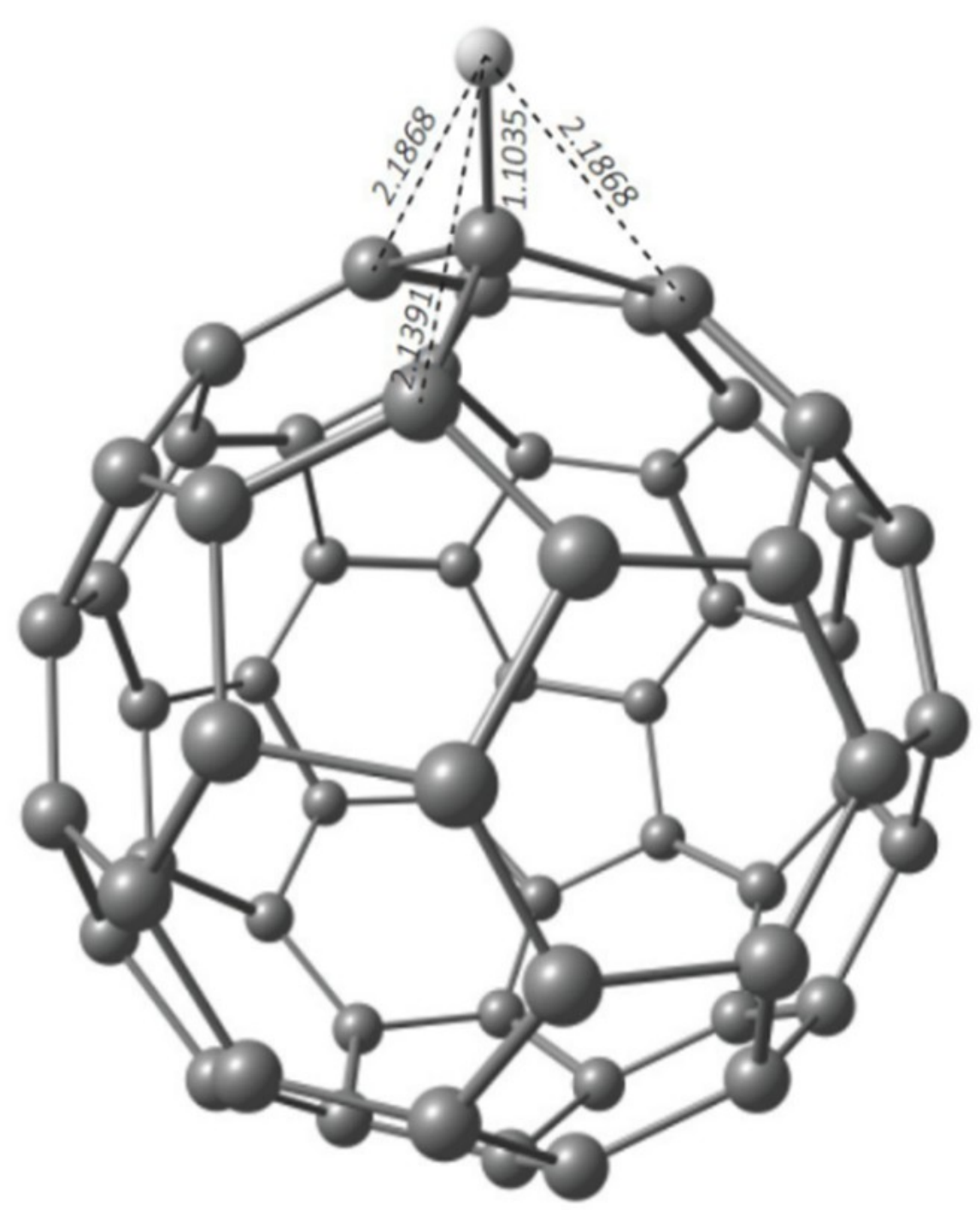

***Figure 1*** *Structure of $C_{60}H^+$ with the C-$H^+$ bond length shown in Angstrom*

The proton binds to the carbon by a covalent bond of 1.1035 Ā̄, distorting the original $sp^2$ configuration to the tetragonal $sp^3$. To understand this transformation, one has to consider that in $C_{60}$ the symmetry axes are not on vertices (on atoms), but in the centers of hexagons and pentagons. Hence, the attachment of a proton to a carbon atom results in a significant symmetry breaking and not simply to a reduction to a symmetric top group. Adding a proton thus leads directly to the asymmetric top symmetry $C_s$. Only a single mirror plane is preserved (see Figure 2) and the emerging dipole becomes slightly misaligned with the C-$H^+$ bond.

Of course, the question emerges whether there are structural isomers depending on which carbon carries the proton. However, since all carbon atoms are symmetrically equivalent due to the isolated pentagon rule, all carbon atoms are situated at an apex of two sidewise joined hexagonal rings enclosing a pentagon, so the proton binding site is immaterial. Singly protonated $C_{60}$ has no distinct structural isomers and there is only one local minimum corresponding to a stable configuration. When quantum mechanical proton hopping (tunneling) occurs between these equivalent sites, spectroscopic features (line broadening or doublet forming) change and it is necessary to use a molecular symmetry group containing elements describing proton tunneling motion.[25,26] In Figure 2, the molecule-fixed Cartesian axes and the electrostatic potential distribution are shown.

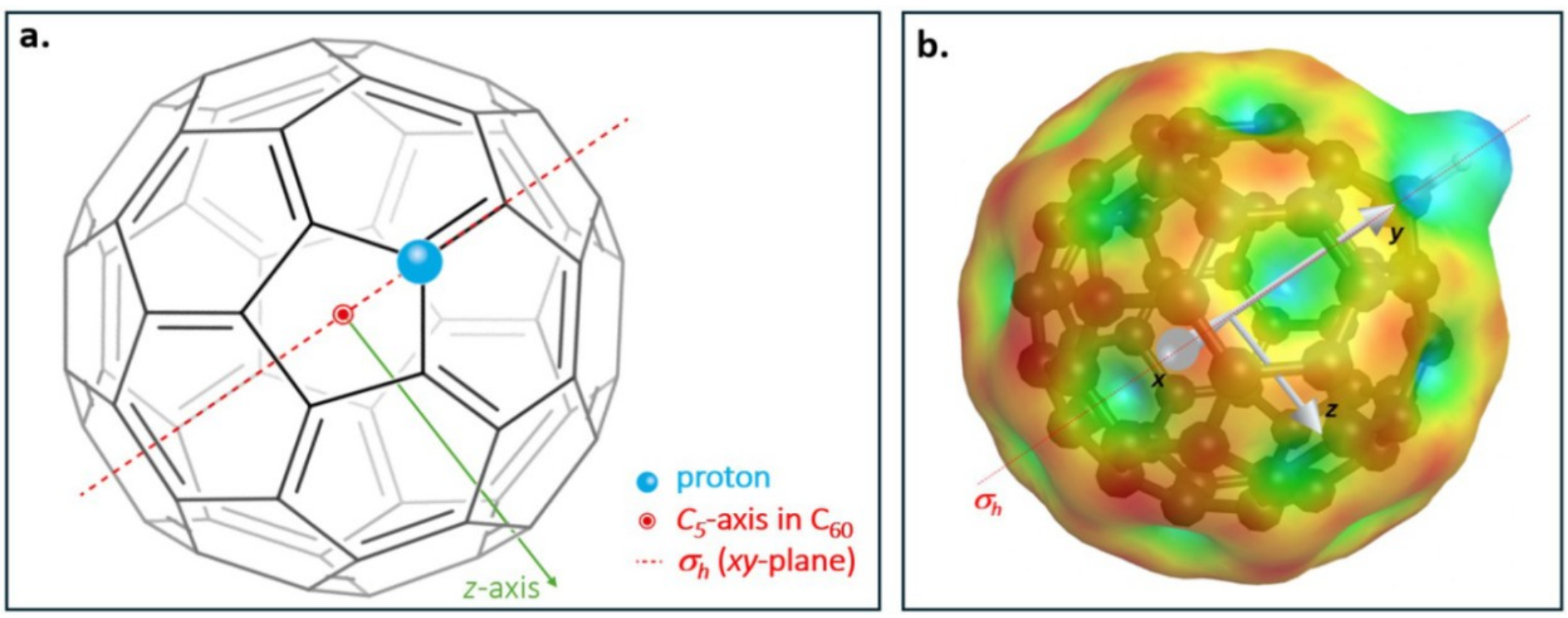

***Figure 2. a.)*** *Sketch of protonated $C_{60}$ showing the molecular mirror plane edge-on, the proton is attached to one of the C-atoms. The mirror plane is the only remaining symmetry element.* ***b.)*** *The electrostatic gradient of the Mulliken charge density of $C_{60}H^+$ was derived from anharmonic Gaussian 16 calculations and rendered in GaussView. The gradient goes from red to blue, from negative to positive. The size of the dipole is 3.8619 D. The molecule-fixed Cartesian axes x, y and z are indicated; the z axis is perpendicular to the xy plane. The bond containing the proton, and the dipole moment vector components are in the xy plane.*

The charge density is distributed over the $C_{60}$ body, but at the proton there is a local density maximum. The dipole arises mainly from the separation between the proton charge and the center of mass of the molecular ion. As mentioned above, there is only one local potential energy minimum corresponding to a stable configuration, as all C-atoms in $C_{60}$ are identical the location of the proton is immaterial. We confirmed this by calculating multiple proton locations using the same C atom coordinates. The results of calculation for the rotational constants and the dipole moments based on the equilibrium $\mathbf{r_e}$ structure as obtained from harmonic calculation using the B3LYP/6-311+G(d,p) level are listed in Table 1.

***Table 1***. *Equilibrium rotational constants and space-fixed Cartesian dipole moment components for three different proton locations from harmonic calculations using the B3LYP/6-311+G(d,p) level. The dipole moment components correspond to the field-independent basis used in Gaussian. The very small difference in $\mu_y$ is probably due to a rounding error. The numbering of the carbons is not shown in Figure 1.*

| Position $C_2$ | Position $C_{35}$ | Position $C_{43}$ |
|---|---|---|
| A: 83.76 MHz | A: 83.76 MHz | A: 83.76 MHz |
| B: 83.47 MHz | B: 83.47 MHz | B: 83.47 MHz |
| C: 83.17 MHz | C: 83.17 MHz | C: 83.17 MHz |
| $\mu_x$: 1.0771 D | $\mu_x$: 1.0771 D | $\mu_x$ : -1.0771 D |
| $\mu_y$: 3.4012 D | $\mu_y$: 3.4011 D | $\mu_y$: 3.4012 D |
| $\mu_z$: 0.0000 D | $\mu_z$:  0.0000 D | $\mu_z$: 0.0000 D |
| $\mu$: 3.5676 D | $\mu$. 3.5676 D | $\mu$: 3.5677 D |

For correct rotational intensities, molecule-fixed Cartesian dipole components have to be used, which are usually called the principal moment axis components. The x,y,z axes in Figure 2 are molecule-fixed Cartesian axes.

It is also important to consider that molecules are not rigid even at absolute zero, since the constituting atoms are in zero-point vibrational motion. Therefore, the zero point average geometry is somewhat different from the equilibrium geometry obtained from harmonic vibrational potentials. For small polyatomic molecules an empirical relationship was obtained between the equilibrium and zero-point average moments of inertia.[27] It predicts about 0.3% difference between the rotational constants of the $r_e$ and $r_0$ structures of small molecules. Instead of using this statistical estimate, anharmonic vibrational calculations were carried out. Both harmonic and anharmonic runs were performed and converged to $C_s$ symmetry (see Section 2.). The comparisons among constants from the harmonic and anharmonic calculations are summarized in Table 2. As anharmonic methods are computationally very expensive for such large molecules, the B3LYP/6-31G level was used instead of the B3LYP/6-311+G(d,p) level used in the harmonic calculations.

***Table 2***. *Comparison of rotational constants (in MHz) and the Ray-type asymmetry parameter $\kappa=(2B-A-C)/(A-C)$ in the $III_r$ axis representation (x,y,z>a,b,c). This shows that $C_{60}H^+$ is a near oblate asymmetric top. These kappa values are small showing that the inertial asymmetry is rather small.*

| Rotational Constants | $r_e$ harmonic | $r_e$ anharmonic | $r_0$ anharmonic | $r_0$-$r_e$ anharmonic difference |
|---|---|---|---|---|
| A | 83.76 | 82.976 | 82.433 | -0.543 |
| B | 83.47 | 82.632 | 82.082 | -0.550 |
| C | 83.17 | 82.388 | 81.848 | -0.540 |
| κ | + 0.01695 | - 0.17007 | - 0.2000 | ---------- |

***Table 3***. *Quartic centrifugal constants (in MHz) for the $III_r$ asymmetric top reduction*

| | |
|---|---|
| $\Delta J$ | $1.7409 \times 10^{-8}$ |
| $\Delta K$ | $-5.1020 \times 10^{-10}$ |
| $\Delta JK$ | $2.4323 \times 10^{-10}$ |
| $\delta J$ | $2.8096 \times 10^{-11}$ |
| $\delta K$ | $-1.2876 \times 10^{-9}$ |

While quartic centrifugal distortion constants were used in the simulations, as taken from the anharmonic calculations, shown in Table 3, they are so small that they may be disregarded at the J values used in this study and this results in the same effective rotational constants. At higher temperatures (and hence higher J values), however, centrifugal distortions may have to be included in rotational simulations. In the following simulations the rigid rotor model was used.

## 4. Simulated rotational spectra of $C_{60}H^+$

In the present simulations, the rotational constants in the fourth column of Table 2 and ground-state vibrational averaged dipole moment components in the principal axis system were used: $\mu_a$= 3.6057 D, $\mu_b$ = -1.383 D and $\mu_c$ = 0 D. The orientations of the **a**, **b** and **c** principal axes and the direction of the dipole moment are shown in Figure 3. The molecular symmetry plane is indicated in light blue color; axes **a** and **b** are in the symmetry plane and axis **c** is perpendicular to it. The dipole moment direction coincides with the C-H bond, while axis **a** is a little rotated away.

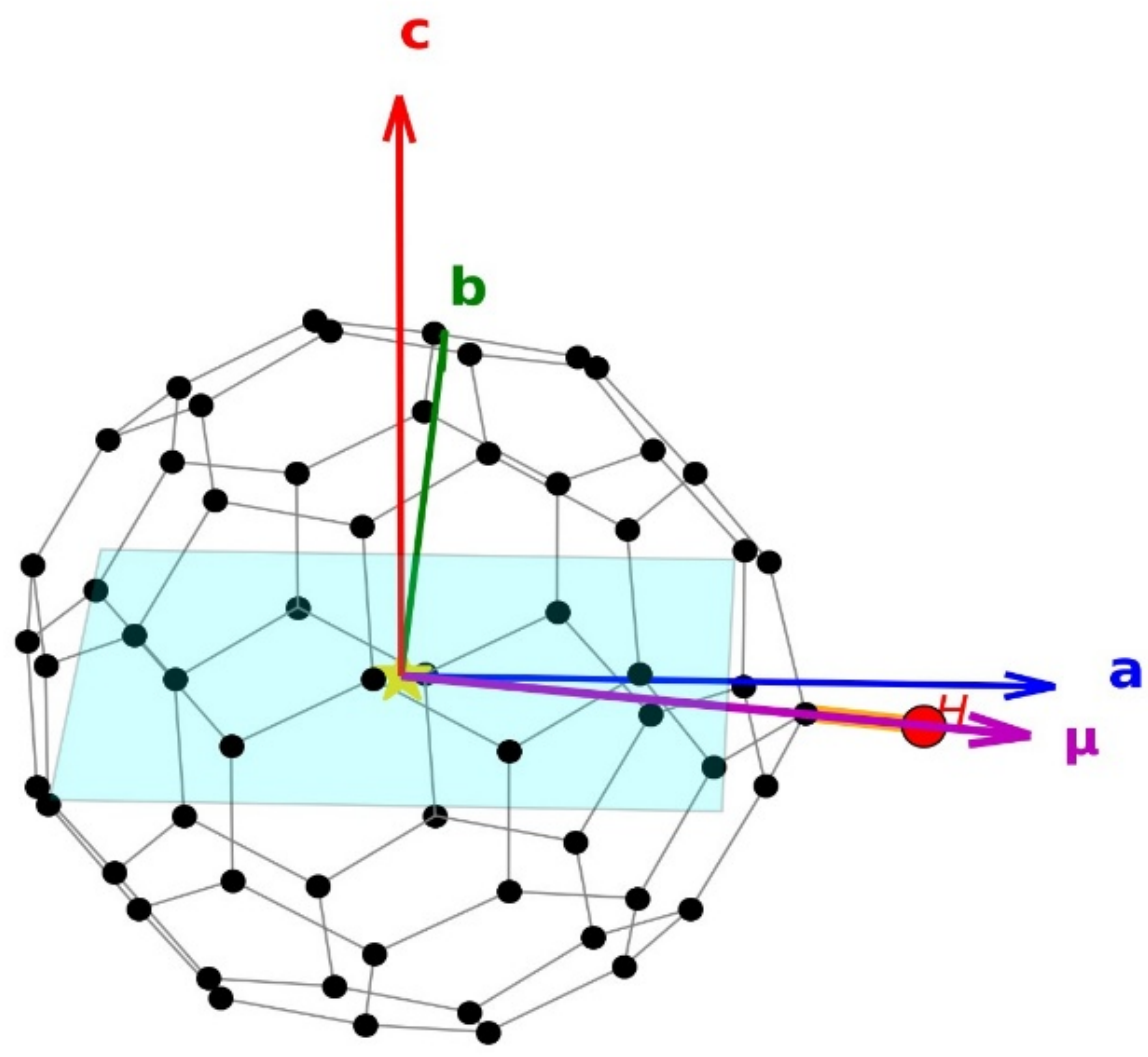

***Figure 3.*** *The position of the principal rotational axes in* $C_{60}H^{+}$.

The positions of the principal rotational axes are not used by the Gaussian anharmonic vibrational calculations, as the positions of the nuclei, dipole components, etc. are referred to a space-fixed static Cartesian system of axes. To obtain the dipole moment components in the molecule-fixed axes, we use the inertial tensor that contains the Eulerian angles, describing the relative position of the two axis systems.[28] In Table 4, these principal axis dipole moment components in the harmonic and anharmonic equilibrium structures and the anharmonic ground-state vibrational average structure are given (for the corresponding rotational constants see the $2^{nd}$ to $4^{th}$ columns in Table 2).

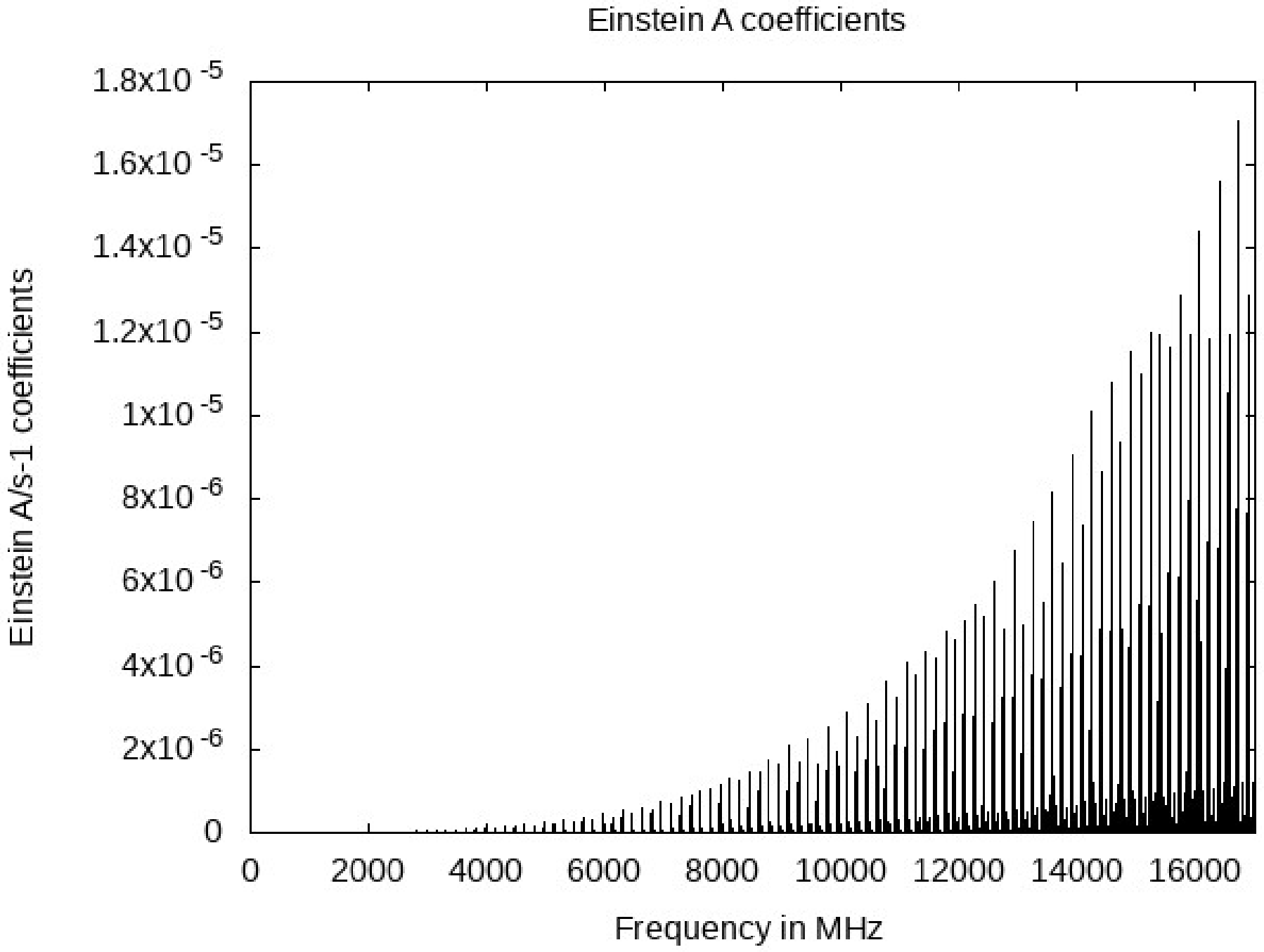

***Figure 4.*** *The Einstein A coefficients.*

The Einstein A coefficients correspond to spontaneous emission rates from the upper rotational levels. The Einstein A coefficient scales with the square of the transition moment in Debye and the cube of the line frequency in MHz units.

***Table 4.*** *The principal axis dipole moment components (in Debye) in the equilibrium and ground vibrational average state structures.*

| Computational models | Total | A axis | B axis | C axis |
|---|---|---|---|---|
| Harm. Equilibrium | 3.5676 | 3.2068 | 1.5636 | 0 |
| Anharm. Equilibrium | 3.8270 | 3.5345 | -1.4674 | 0 |
| Anharm. ground vibr. aver. | 3.8619 | 3.6057 | -1.3830 | 0 |

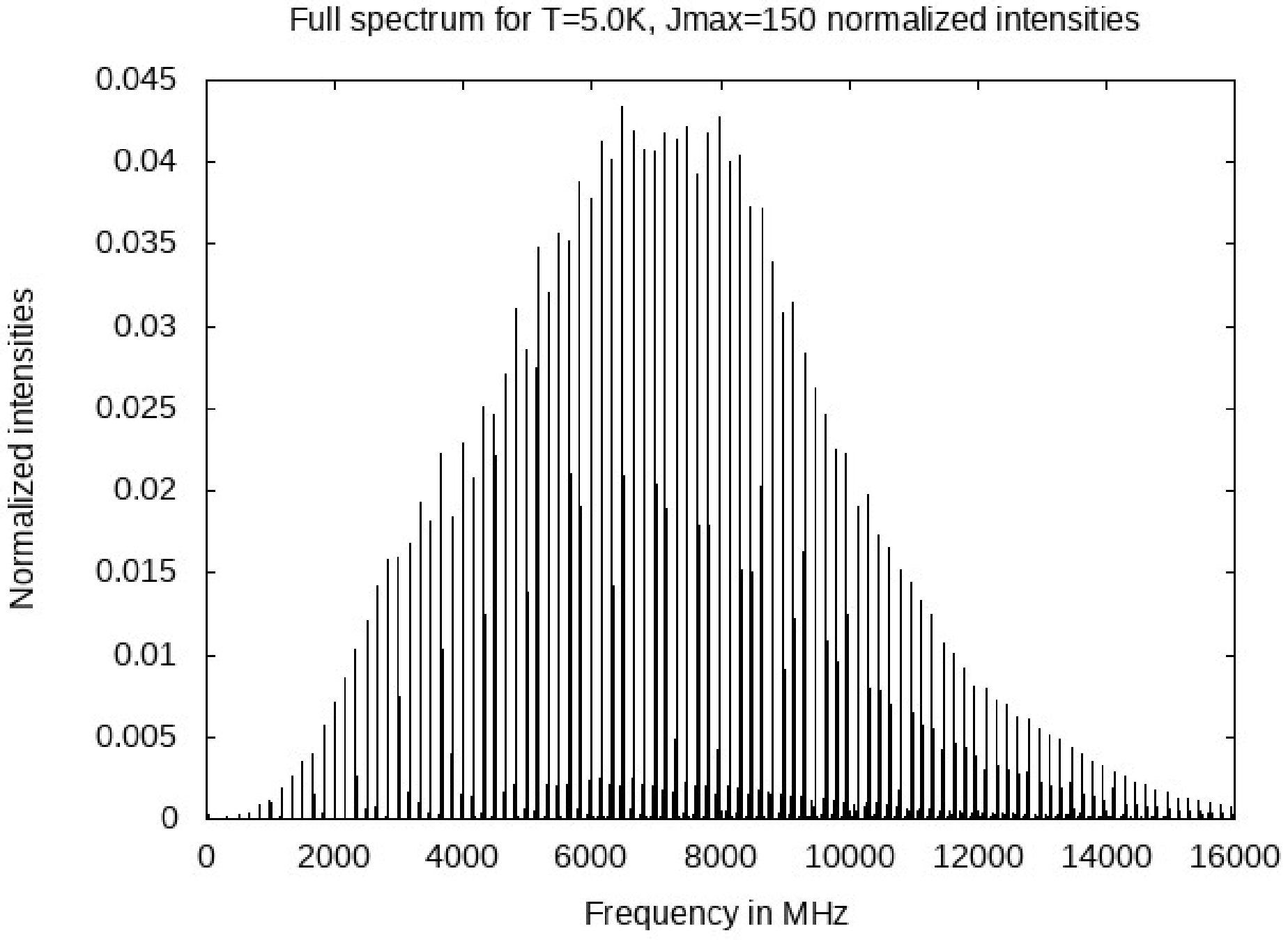

***Figure 5*** *Simulated rotational spectrum of* $C_{60}H^+$ *at T=5 K.*

Figures 5–8 show the predicted spectra in normalized intensities, which are the default in PGOPHER and which contain rotational partition functions and Boltzmannian level populations. An excitation temperature of 5 K was chosen in Figure 5 to resemble laboratory molecular beam microwave spectroscopy and dark molecular cloud conditions and applies for thermal equilibrium. The dipole moment components used are those in the last row of Table 4. In the simulations, individual lines were given zero Gaussian and Lorentzian line widths and no Doppler width was used. The dense spectrum contains a large number of spectral lines, merging together at the low resolution of the plot and due to the relatively small number points used in the simulation. Thus, the intensity fluctuations in Figures 5 and 6 are artifacts and do not correspond to the true rotational structure. The purpose of showing these simulations is to give the frequency range and its shift upon increasing the temperature.

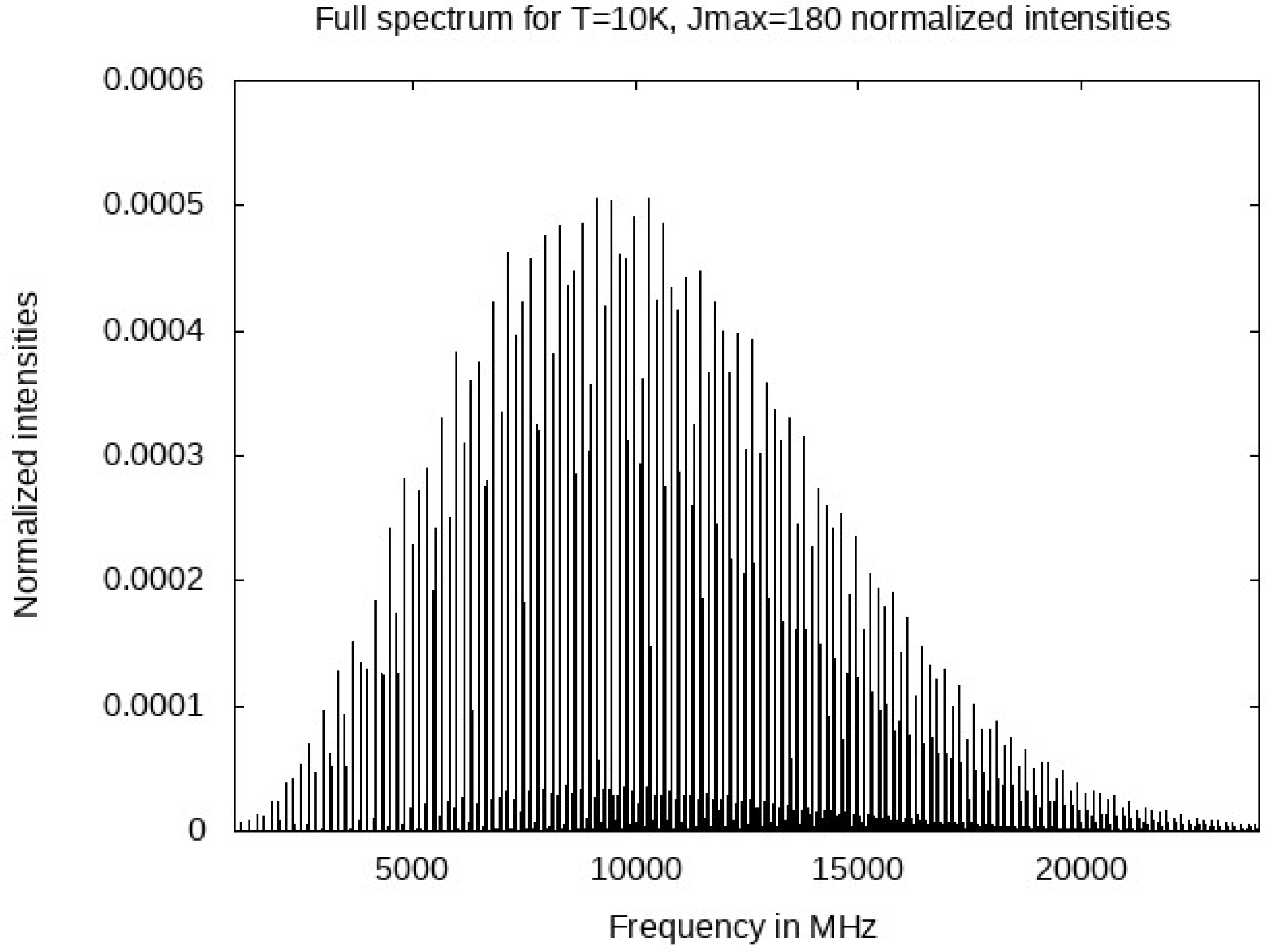

***Figure 6***. *Simulated rotational spectrum at T = 10 K.*

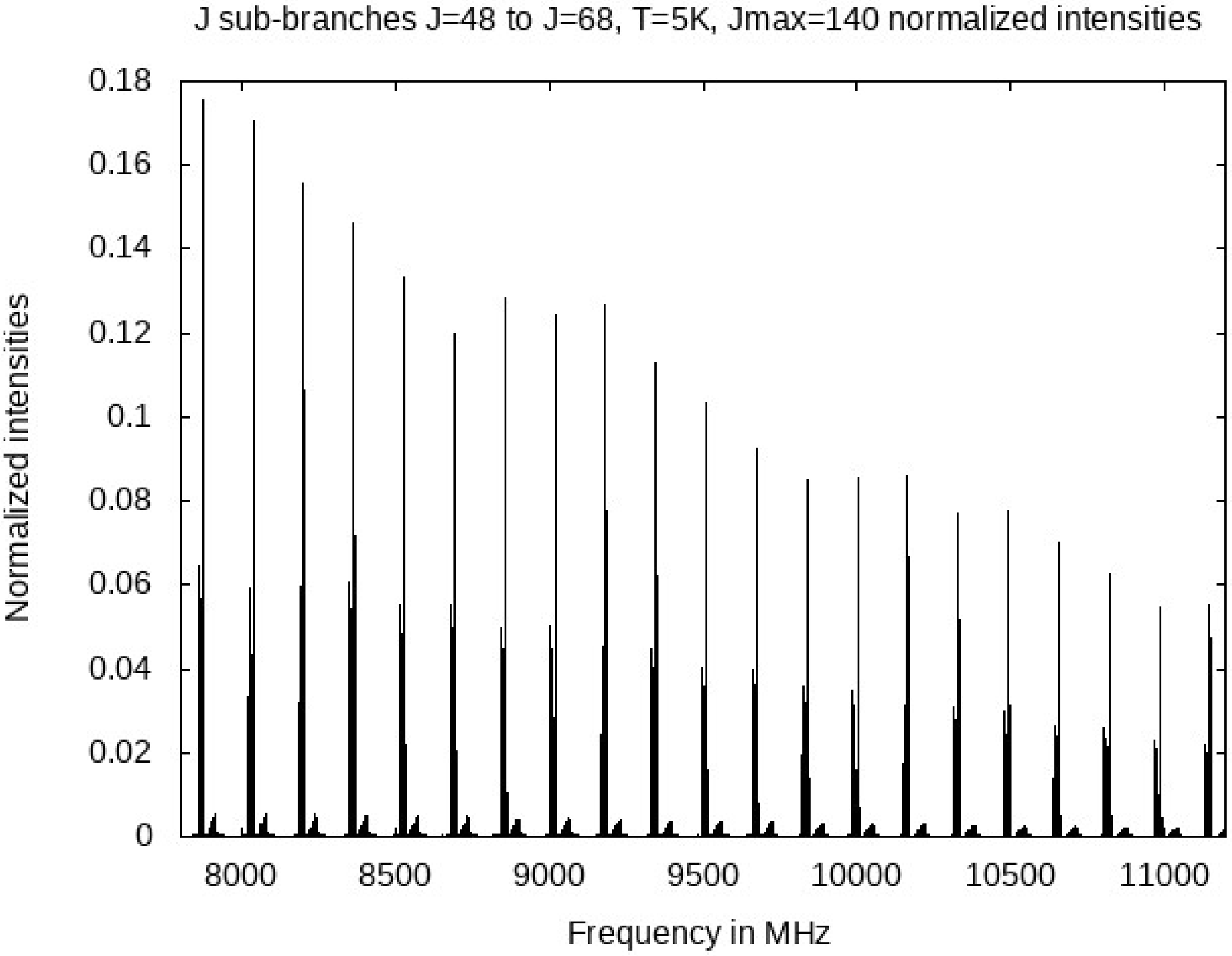

***Figure 7.*** *A series of J sub-branches at T=5 K. These sub-branches lie in the X band of the Greenbank Radio Telescope. Each sub-branch contains a dense K structure. The separation between the J sub-branches is (B+C), which is about 164 MHz. The K structure in the J=60 J-subbranch is shown in Figure 8.*

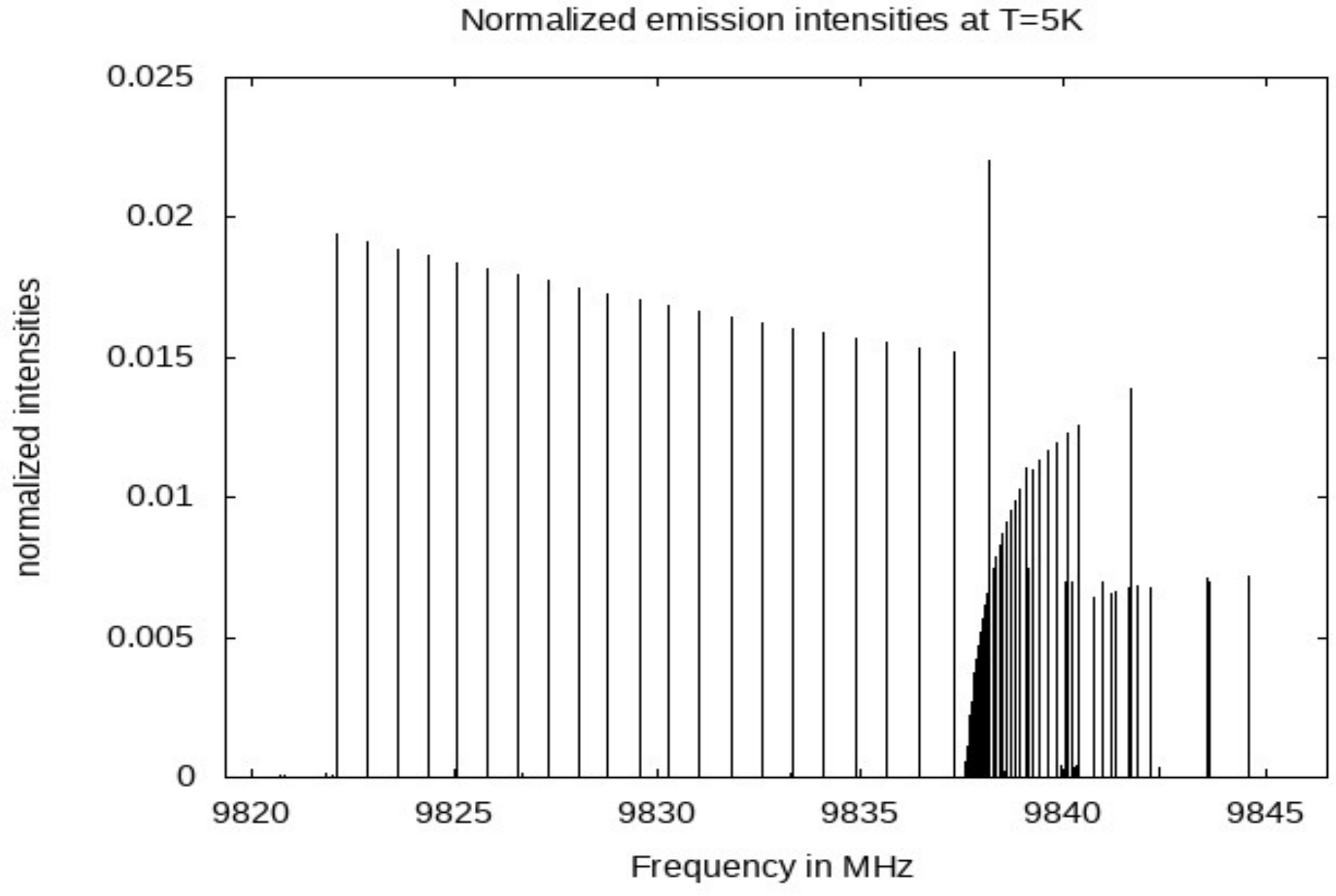

***Figure 8.*** *Detail of the emission spectrum of the J=60 sub-branch for T=5 K.*

Figure 8 shows an enlarged portion of the J=60 sub-branch, which contains only a-type transitions. Most of these form a regular, comb-like arrangement. The separation between the lines in the comb-like part of the spectrum is less than 1 kHz. In asymmetrical $C_s$ rotors the rotation transition symmetry is A" or A' .

In Table 5 representative values of the rotational partition sums for $C_{60}H^+$ are listed for temperatures of 5, 10, 15 and 20 Kelvin, as calculated in the symmetric top approximation. The rotational constants from the second column in Table 2 are: $B_1$= A = 83.76 MHz, $B_2 = \sqrt{(B*C)}$ = 83.2934084 MHz, where $B_1$ and $B_2$ are the rotational constants of the one-dimensional and two-dimensional rotors of the symmetric top.[29]

***Table 5.*** *Rotational partition sums for four different temperatures calculated using the symmetric top approximation*

| *Temperature (K)* | $Q_{rot}$ |
|---|---|
| 5.0 | 78191 |
| 10.0 | 217155 |
| 15.0 | 372926 |
| 20.0 | 519159 |

**5. Discussion**

The present work aims at recommending efforts in laboratory microwave spectroscopy and radio astronomy for finding protonated $C_{60}$ in cosmic sources, thus offering detection possibilities in addition to infrared astronomical methods, when laboratory IR spectroscopy and eventually optical spectroscopy of the visible spectrum of $C_{60}^+$ will be known[21,22]. Specifically, rotational spectroscopy offers the opportunity to detect $C_{60}$ derivatives in regions devoid of UV irradiation, such as dark cloud cores. This can then be used to infer the presence of $C_{60}$ itself and its abundance can be derived indirectly, as is done for the cyanogen derivatives of small PAHs.[15,17]

As emphasized in the Introduction, the absence of UV in dark cloud cores inhibits the use of the commonly used IR fluorescence detection methods. Likewise, the high dust extinction at visible and far-red wavelengths through such cores hampers the detection of electronic transitions in the spectra of background stars. Chemical studies of neutral and cationic $C_{60}$ have been developed describing the formation of $C_{60}H^+$ and other hydrogenated fullerene cations.[30,31]

Following the unpublished opinion of Nobelist Harry Kroto, it is expected that the most abundant hydrogenated buckminsterfullerene compound in the diffuse interstellar medium is $C_{60}H^+$.

The radio-astronomical search for polar fullerene forms and derivatives is undoubtedly going to be more difficult than the search for small molecules, because the large rotational partition sums result in a large number of weak lines. This obviously is the case for $C_{60}H^+$. Table 4 contains a sample of rotational partition sums for very low temperatures. The population of the rotational levels increases quickly at elevated temperatures, so that the spectrum becomes densely filled with lines. At temperatures around 100 K and above, the spectra may even become a weak continuum. The comparison of Figures 5 and 6 demonstrates the shift of the most intense lines from about 7 GHz to about 10 GHz upon temperature increase from 5 K to 10 K. Radio astronomical searches for $C_{60}H^+$ are expected to be the most successful in the coldest cosmic locations. In the cores of dark dense molecular clouds (such as the TMC1 Taurus molecular cloud), interstellar dust grains in the coldest, densest part of the cloud extinguish visible and infrared radiation, so that molecular detection is only possible by radio astronomy. In photo-dissociation regions (PDRs), a star irradiates the surrounding neutral gas with far-ultraviolet photons. Rather than collisional equilibrium, in PDRs, the rotational population is set by UV excitation followed by ro-vibrational cascade. The resulting rotational population can be well approximated by a distribution at an effective rotational temperature of $T_{rot} = hc\nu_{IR}/6k$ with $\nu_{IR}$ the average vibrational frequency and the factor 6 takes approximately care of the sum over the K-ladders.[32] With $\nu_{IR} \approx 600$ cm$^{-1}$, we find $T_{rot} \approx 150$

K. Thus, the rotational excitation and partition function will be large, causing the rotational spectrum to become very weak and unfavorable for detection by radio astronomy.

The question emerges whether the simulations presented in this work are applicable, considering that these assume thermal equilibrium. This question may be answered with the computed Einstein A coefficients in Figure 4. Because of the larger dipole moment and Einstein A coefficients, collisional de-population of $C_{60}H^+$ rotational levels should occur faster than for the neutral radical species. This de-population factor may be estimated by the so-called critical density term $n_{cr}$ given by the ratio of the Einstein A coefficient and the collisional de-excitation coefficient, γ

$$n_{cr} = \frac{\sum_{l< u} A_{u,l}}{\sum_{l\neq u} \gamma_{u,l}}$$

where the u,l indices refer to the upper and lower levels, respectively (See chapter 4 in Ref. 32). At densities larger than $n_{cr}$ thermal equilibrium is established. For ion-neutral interactions, the collisional de-excitation coefficient, γ, is given by the Langevin rate, $\approx 5 \cdot 10^{-8}\ cm^3/s$ for $H_2$.[32] From Figure 4, we see that the largest Einstein A coefficient is around $1 \cdot 10^{-5}\ s^{-1}$ and the critical density is about 200 $cm^{-3}$. Typical densities are about 10000 $cm^{-3}$, so that collisions are much faster than radiative de-excitation and therefore thermal equilibrium is established. This result supports the thermal equilibrium model for spectral simulations using the PGOPHER software. Thermal equilibrium is involved in the plots displayed in Figures 5-8.

As to the possibility of emission from protonated fullerenes and similar polar carbon molecules contributing to the anomalous microwave emission (AME[33]), as proposed by Iglesias-Groth,[34] high temperature rotational simulations are needed but not yet available. The presently used 5 K excitation temperature is chosen on the basis of the work of Wenzel *et al.* on cyanocoronene[17] and it is appropriate for the dark cloud core TMC1.

Carefully inspecting the values in Table 2, one notices another important aspect of the quantum-chemical calculations performed for $C_{60}H^+$, which is a near oblate asymmetric top, essentially an almost spherical top. The DFT-computed values for the rotational constants obtained in harmonic and anharmonic calculations are rather different, so that computing rotational spectra in the harmonic approximation may not be sufficiently accurate for radio astronomical searches. Considering that the J sub-branch separation is given by (B+C), the sub-branch separations are sensitive to the value of the rotational constants.

In astronomical studies of the rotational emission of interstellar molecules, intensities are generally given in units of K km/s. Using the intensities provided by PGOPHER and taking $I_{rot}$ in units of Watts/molecule, astronomical intensities have been calculated as

$$T\Delta v = \frac{c^3}{8\pi v^3 k} I_{rot} N \approx 7.8 \times 10^{-4} \left(\frac{10\ GHz}{v}\right)^3 \left(\frac{I_{rot}}{10^{-35} \frac{W}{molecule}}\right) \left(\frac{N_{tot}}{10^{14} \frac{molecules}{cm^2}}\right) K\ km/s,$$

with $N_{tot}$ the total number of molecules along the line of sight. The values in Table 6 have been calculated for a kinetic temperature of 5 K and total $C_{60}H^+$ column densities of $10^{13}$ and $10^{14}$ $cm^{-2}$, which is reasonable given the excitation temperature of 5 K and column density of $3 \times 10^{12}$ $cm^{-2}$ observed for cyanocoronene.[17] As intensities depend linearly on the adopted column density, the intensities provided in Table 6 are easily extrapolated to other column densities. For comparison, we estimate a 3σ-detection limit of about $10^{-3}$ K km/s from the spectra of the rotational transitions of cyanocoronene in this same frequency range.[17] This sensitivity can be greatly improved by stacking of lines, as for random noise the sensitivity improves as the square root of the number of lines stacked. In addition, special filtering techniques have been developed to improve the detectability of rotational transitions.[15,35]

It appears that $C_{60}H^+$ has the right rotational structure for detection using the rotational comb method reinforced by the spectral match filtering. Figure 7 shows a J comb with tooth separation of about 167 MHz Each tooth contains the K structure in its regular part as a finer comb with tooth separation of about 7.5 KHz. Thus, the rotational comb for each J+1 → J transition has teeth as a series of finer K combs.

To detect $C_{60}H^+$ in the TMC1 molecular black cloud at T = 5 K, the Green bank Telescope (GBT) could be used in the X frequency band (8.0 – 11.6 GHz) using the VEGAS astronomical spectrometer that has sensitivity down to 0.7 kHz.

**Table 6:** Intensities at T=5 K of selected lines near 9.8 GHz in units of K km/s for adopted total column densities of $10^{13}$ and $10^{14}$ $cm^{-2}$.

| frequency | Intensity | $E_u$ | $E_l$ | $T_{mb}*\Delta v$ | |
|---|---|---|---|---|---|
| MHz | W/molecule | MHz | MHz | K km/s | |
| | | | | $N=10^{14}$ $cm^{-2}$ | $N=10^{13}$ $cm^{-2}$ |
| 9822.13 | 3.92E-35 | 299585.859 | 289763.7291 | 3.22E-03 | 3.22E-04 |
| 9822.87 | 5.57E-36 | 299629.806 | 289806.9355 | 4.57E-04 | 4.57E-05 |
| 9823.6102 | 3.84E-35 | 299672.926 | 289849.3161 | 3.15E-03 | 3.15E-04 |
| 9824.3505 | 3.80E-35 | 299715.216 | 289890.8657 | 3.12E-03 | 3.12E-04 |
| 9825.0913 | 3.76E-35 | 299756.67 | 289931.5787 | 3.09E-03 | 3.09E-04 |
| 9825.8325 | 3.72E-35 | 299797.281 | 289971.4489 | 3.05E-03 | 3.05E-04 |
| 9826.5743 | 3.69E-35 | 299837.044 | 290010.4692 | 3.03E-03 | 3.03E-04 |
| 9827.3171 | 3.65E-35 | 299875.949 | 290048.632 | 2.99E-03 | 2.99E-04 |
| 9828.061 | 3.61E-35 | 299913.99 | 290085.9286 | 2.96E-03 | 2.96E-04 |
| 9828.8064 | 3.58E-35 | 299951.155 | 290122.349 | 2.94E-03 | 2.94E-04 |
| 9829.5536 | 3.55E-35 | 299987.436 | 290157.8824 | 2.91E-03 | 2.91E-04 |
| 9831.0556 | 3.48E-35 | 300057.291 | 290226.2352 | 2.85E-03 | 2.85E-04 |
| 9831.8118 | 3.45E-35 | 300090.835 | 290259.0236 | 2.83E-03 | 2.83E-04 |
| 9832.5725 | 3.42E-35 | 300123.434 | 290290.8616 | 2.80E-03 | 2.80E-04 |
| 9833.3392 | 3.40E-35 | 300155.066 | 290321.7263 | 2.78E-03 | 2.78E-04 |
| 9834.1133 | 3.37E-35 | 300185.704 | 290351.5905 | 2.76E-03 | 2.76E-04 |
| 9834.8971 | 3.35E-35 | 300215.319 | 290380.4215 | 2.74E-03 | 2.74E-04 |
| 9835.6936 | 3.32E-35 | 300243.872 | 290408.1788 | 2.72E-03 | 2.72E-04 |
| 9836.507 | 3.30E-35 | 300271.319 | 290434.8118 | 2.70E-03 | 2.70E-04 |
| 9837.3436 | 3.29E-35 | 300297.599 | 290460.255 | 2.69E-03 | 2.69E-04 |
| 9837.344 | 3.29E-35 | 300297.598 | 290460.2544 | 2.69E-03 | 2.69E-04 |
| 9837.7536 | 6.33E-36 | 301417.538 | 291579.7848 | 5.18E-04 | 5.18E-05 |

## Conclusions

We present predicted pure rotational spectra for protonated $C_{60}$ at different temperatures. $C_{60}H^+$ is a near oblate asymmetric top of $C_s$ symmetry, being very close to spherical geometry, which gives the species its peculiar spectrum. Although rotational constants in the molecule-fixed frame were carefully derived from DFT-optimized geometries using harmonic and anharmonic approaches, line positions are not sufficiently accurate to directly allow for a search of astronomical spectra. Nonetheless, we believe that the present results provide new insights into the rotational structure of nearly symmetric polar fullerene derivatives and may guide future laboratory and astronomical studies.

## Acknowledgements

J.O. thanks the Nederlandse Organisatie voor Wetenschappelijk Onderzoek (NWO) and the SurfSARA Supercomputer Centre for providing computational time and resources (grant no. 2024.009). V.J.E. acknowledges support from the Internal Scientist Funding Model (ISFM) Laboratory Astrophysics Directed Work Package at NASA Ames. Computer time from the Aiken cluster of the NASA Advanced Supercomputer (NAS) is gratefully acknowledged.

## Conflicts of interest

The authors declare no competing financial interest.